\documentclass[aps,prl,epsfigure,twocolumn,superscriptaddress]{revtex4-2}
\usepackage[colorlinks=true,linkcolor=blue,urlcolor=blue,citecolor=blue,pdfusetitle]{hyperref}

\usepackage{graphicx}
\usepackage{dcolumn}
\usepackage{bm,upgreek}
\usepackage{soul}
\usepackage[normalem]{ulem}
\usepackage{printlen}
\uselengthunit{in}
\usepackage{color}

\usepackage{physics}
\usepackage{hyperref}
\usepackage{bibunits}

\begin{document}

\newcommand{\papertitle}{Collective resonance displacement in strongly driven cold atoms}

\title{ \papertitle}

\author{Mateus A. F. Biscassi}
\email{mateus.af.biscassi@gmail.com}
\affiliation{Universit\'e C\^ote d'Azur, CNRS, INPHYNI, France}
\affiliation{Departamento de F\'{\i}sica, Universidade Federal de S\~{a}o Carlos, Rodovia Washington Lu\'{\i}s, km 235 - SP-310, 13565-905 S\~{a}o Carlos, SP, Brazil}

\author{Robin Kaiser}
\email{robin.kaiser@univ-cotedazur.fr}
\affiliation{Universit\'e C\^ote d'Azur, CNRS, INPHYNI, France}

\author{Mathilde Hugbart}
\email{mathilde.hugbart@univ-cotedazur.fr}
\affiliation{Universit\'e C\^ote d'Azur, CNRS, INPHYNI, France}

\author{Romain Bachelard}
\email{romain@ufscar.br}
\affiliation{Departamento de F\'{\i}sica, Universidade Federal de S\~{a}o Carlos, Rodovia Washington Lu\'{\i}s, km 235 - SP-310, 13565-905 S\~{a}o Carlos, SP, Brazil}

\date{\today}

\begin{abstract}
Cold atoms are promising platforms for metrology and quantum computation, yet their many-body dynamics remains largely unexplored. We here investigate Rabi oscillations from optically-thick cold clouds, driven by high-intensity coherent light. A dynamical displacement from the atomic resonance is predicted, which can be detected through the collective Rabi oscillations of the atomic ensemble. Different from linear-optics shifts, this dynamical displacement grows quadratically with the optical depth, yet it reduces with increasing pump power as dipole-dipole interactions are less effective.
\end{abstract}

\maketitle

\section{Introduction} Cold atom setups are important contenders for quantum simulations and computation, or metrology~\cite{Johanning2009,Bloch2012,Gross2017,Chang2004}. The advent of quantum microscopy allows one to probe individually the emitters, which is important to witness the emergence of correlations and entanglement~\cite{Kuhr2016}. Interestingly, while light is used to interrogate the atoms, it also induces interactions between them, as the radiation of emitters into common electromagnetic modes leads to an effective dipole-dipole interaction~\cite{Lehmberg1970}. This coupling is particularly relevant in optical atomic clocks, where it leads to cooperative shifts~\cite{Cidrim2021,Ross2024}.

While cooperative effects were first discussed in the context of fully-inverted systems and their decay toward the ground state, exploring multiple-excitation, superradiant Dicke states~\cite{Dicke1954}, it has recently attracted a renewed attention as superradiance and (long-lived) subradiant states were observed close to the ground state~\cite{Guerin2016, Araujo2016,Roof2016,Cipris2021}. In this regime, where the system exhibits a linear response and can be described with a classical approach~\cite{Svidzinsky2010,Cottier2018}, collective shifts have been reported~\cite{Friedberg1973,Javanainen2014, Jenkins2016, Jennewein2016, Peyrot2018, Guerin2019, Glicenstein2020, Kemp2020,hofer2024,Hsu2024}, which arise from the Hamiltonian terms of the dipole-dipole interaction.


However, the existence of shifts remains largely unexplored in the many-excitation regime, where the near-field Hamiltonian terms have mainly been pointed out to induce a broadening detrimental to collective decay in the cascade configuration~\cite{Gross1982}. In presence of a strong drive, atomic coherences are reduced, thus limiting the effectiveness of dipolar couplings. This raises the question of whether collective shifts may arise in this regime, and lead to extra contributions, especially in the context of atomic clocks where dense samples and strong pulses drive the system far from the ground state~\cite{Chang2004, Cidrim2021}.

\begin{figure}[htpb]
    \centering
    \includegraphics[width=\columnwidth]{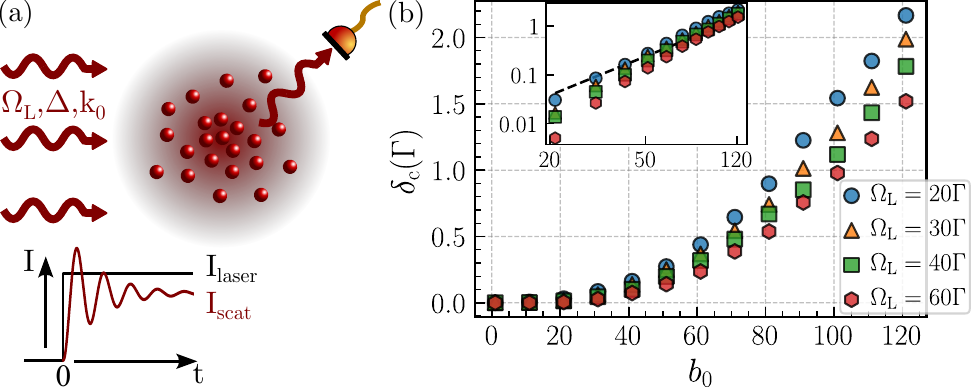}
    \caption{(a) Schematic of the three-dimensional cloud driven with a monochromatic coherent light starting from time $t=0$ ($I_\mathrm{laser}$ in the intensity graph), which makes the atoms undergo Rabi oscillations ($I_\mathrm{scat}$ in the same graph). The driving field is characterized by the driving strength $\Omega_\mathrm{L}$, detuning $\Delta$ and wavevector $\mathrm{k}_0$. (b) Collective dynamical resonance displacement $\delta_\mathrm{c}$ as a function of optical depth $b_0$, for different driving Rabi frequency $\Omega_\mathrm{L}$. The inset shows the same data in log-log scale, showing that the displacement scales quadratically with $b_0$, but decreases with increasing drive strength due to reduced atomic coherences and weakened dipole-dipole interactions.}
    \label{fig:schemeAndResult}
\end{figure}

In this work we investigate theoretically the dynamics of strongly driven, optically thick atomic clouds, and report the emergence of a collective dynamical shift in the Rabi oscillation frequency. The coupled-dipole dynamics of three-dimensional clouds are simulated using a mean-field approach, with the atoms abruptly driven far from the ground state. Unlike the linear, weak-drive regime, the resonance displacement scales quadratically with the optical depth, and vanishes in the steady-state regime. Our work thus reveals an out-of-equilibrium collective resonance displacement for strongly-driven atoms, which decays only slowly with the increasing Rabi frequency of the pump (see Fig.\ref{fig:schemeAndResult}). Throughout this work, we focus on dilute disordered clouds, which typically corresponds to the experimental conditions of cold atoms in magneto-optical traps.

\section{Strongly driven coupled atoms in the semi-classical approach} Let us consider a cloud of $N$ two-level atoms driven by a monochromatic laser, whose dynamics is described by a master equation, $\dot{\hat{\rho}}=-i[\hat{H},\hat{\rho}]+\mathcal{L}(\hat{\rho})$, with 
$\hat{\rho}$ the density matrix and setting $\hbar = 1$. The associated Hamiltonian and Lindbladian read
\begin{align}
    \hat{H} =& -\Delta\sum_n\hat{\sigma}^+_n\hat{\sigma}^-_n+\sum_{n,m\neq n}g_{nm}\hat{\sigma}^+_n\hat{\sigma}^-_m\label{eq:DDHamiltonian}
    \\ &+\frac{1}{2}\sum_n(\Omega_\mathrm{L} e^{i\mathbf{k}_0\cdot \mathbf{r}_n}\hat{\sigma}^+_n+\Omega_\mathrm{L} e^{-i\mathbf{k}_0\cdot \mathbf{r}_n}\hat{\sigma}^-_n) \nonumber
    \\ \mathcal{L}(\hat{\rho}) =& \sum_{n,m}f_{nm}(\hat{\sigma}^-_n\hat{\rho}\hat{\sigma}^+_m-\{\hat{\sigma}^+_m\hat{\sigma}^-_n,\hat{\rho}\}),
\end{align}
where $\Delta=\omega_\mathrm{L}-\omega_\mathrm{a}$ is the detuning between the laser frequency and the atomic transition frequency, $\hat{\sigma}^+_n$ ($\hat{\sigma}^-_n$) the raising (lowering) operator for the $n$th atom and $\mathbf{r}_n$ its position. A plane wave with wavevector $\mathbf{k}_0$ and on-resonance Rabi frequency $\Omega_\mathrm{L}$ is considered. In this work, we aim to study dilute disordered clouds, where polarization effects and near-field terms are expected to be negligible. To reproduce such low-density systems using a limited number of atoms while still reaching high optical depths, we adopt the scalar Green’s function
\begin{equation}
    G^{\mathrm{s}}_{nm} = \frac{\Gamma}{2}\frac{e^{ikr_{nm}}}{ikr_{nm}},
\end{equation}
where $r_{nm}$ is the distance between atoms $n$ and $m$ and $k=\omega_\mathrm{a}/c$, with $c$ the speed of light. From the Green's function the coherent $g_{nm} = \mathrm{Im}[G^{\mathrm{s}}_{nm}]$ and dissipative $f_{n\neq m} = \mathrm{Re}[G^{\mathrm{s}}_{nm}]$ couplings are derived, with $\Gamma$ the single-atom spontaneous decay rate and $f_{nn}=\Gamma$.

We consider clouds of $N$ atoms with Gaussian density profile of root-mean-square radius $R$, and on-resonance optical depth $b_0=2N/(kR)^2$. Initially in the ground state, the atoms are suddenly driven, at $t=0$, by the laser and start Rabi oscillations, see Figure~\ref{fig:schemeAndResult}(a). We hereafter focus on \textit{strongly driven} atoms, such that the saturation parameter $s=2\Omega_\mathrm{L}^2/(\Gamma^2+4\Delta^2)$ is large.

Hence, differently from weak-excitation approaches~\cite{Jennewein2016, Peyrot2018, Guerin2019,Kemp2020, Glicenstein2020}, the excited population of the atoms must be accounted for. In order to investigate large three-dimensional systems, yet keeping the system numerically tractable, we resort to a mean-field approximation~\cite{EspiritoSanto2020,CMaximoCooperative,KramerMeanField}: The system is assumed to remain in a product state, $\hat{\rho} = \bigotimes_n \hat{\rho}_n$, thus neglecting interatomic entanglement or correlations, and describing the dynamics using single-atom density matrices only. This approach consists in replacing the field generated by the atoms by its average (that is, its expected value). The scattering dynamics is then described through the expectation values of the single-atom operators $\langle \hat{\sigma}_n^\pm\rangle$ and $\langle \hat{\sigma}_n^z\rangle$.

Each single-atom density matrix $\hat{\rho}_n$ is thus parameterized as
    $\hat{\rho}_n = \frac{1}{2}(\mathbf{1}+2\beta_n^*\hat{\sigma}_n^-+2\beta_n\hat{\sigma}_n^+ +z_n\hat{\sigma}_n^z)$, 
with $\beta_n = \langle \hat{\sigma}^-_n\rangle$ the atomic coherence, $z_n = \langle\hat{\sigma}^z_n\rangle$ the population inversion and $\hat{\sigma}^z_n$ inversion population operator of atom $n$. Inserting the product state into the master equation leads to a set of mean-field coupled equations:
\begin{align}
    \dot{\beta}_n &= \left(i\Delta-\frac{\Gamma}{2}\right)\beta_n+i\Omega_n z_n,\label{eq:bn} \\ 
    \dot{z}_n &= -\Gamma(1+z_n)-4\mathrm{Im}(\beta_n\Omega^*_n),\label{eq:zn}
\end{align}
where $\Omega_n$ is the \textit{local field} acting on atom $n$, composed of both the driving laser and the radiation from the other atoms:
\begin{equation}
    \Omega_n = \frac{\Omega_\mathrm{L} e^{i\mathbf{k\cdot r_n}}}{2}-i\sum_{m\neq n}G^s_{nm}\beta_m.
    \label{eq:localfieldWn}
\end{equation}
From the expression of normalized electric field operator in a direction $\hat{n}$ of the far field, $\hat{E}^+_{\hat{n}} =\sum_n e^{-ik\hat{n}\cdot \mathbf{r}_n}\hat{\sigma}_n^-$, the scattered intensity can be decomposed into elastic and inelastic components $I^{\mathrm{total}}=\langle \hat{E}_{\hat{n}}^- \hat{E}_{\hat{n}}^+\rangle=I_{\hat{n}}^\mathrm{el} +I^\mathrm{inel}$, which here read
\begin{align}
    I_{\hat{n}}^\mathrm{el}&=\left|\sum_{m} e^{-ik\hat{n}\cdot \mathbf{r}_m}\beta_m\right|^2,
    \\ I^\mathrm{inel}&= \sum_m\frac{1+z_m}{2}-|\beta_m|^2.
\end{align}
Hence the dynamics of the excited population, which dominates the scattering in the strong-drive regime, manifests through the inelastically scattered light.

In order to obtain statistically reliable results, the scattered intensity is averaged over 20 disorder realizations. The oscillation frequency is then extracted from the Fourier transform of the intensity signal between $t=0$ and $t=10/\Gamma$, followed by a Lorentzian fit of the resulting spectrum to identify the central frequency. The collective resonance displacement is thus analyzed over a short timescale. In this early-time regime, we assume that recoil effects~\cite{Sesko1991,Ellinger1994} due to the scattering of many incident photons by each atom have a negligible effect, as the absorption is limited to one recoil per atomic transition lifetime.

On the one hand, this semi-classical model captures the weak-excitation approach~\cite{Svidzinsky2010,Courteille2010,Bienaime2012} (``coupled dipole dynamics'') in the $s\to 0$ limit, which is here equivalent to setting $z_n = -1$ in Eqs.~(\ref{eq:bn}, \ref{eq:zn}). On the other hand, the direct interaction through inelastically scattered photons is drastically simplified in the mean-field approximation, since the exchanged field is encapsulated into the average field $\beta_m$ in Eq.~\eqref{eq:localfieldWn}.

\section{Collective Rabi oscillations from the inelastic scattering of strongly driven atoms}

\begin{figure}[htpb]
    \centering
    \includegraphics[width=\columnwidth]{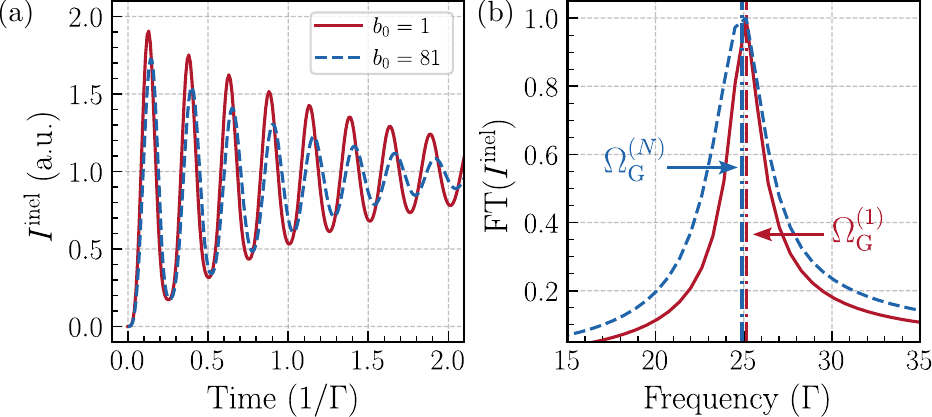}
    \caption{
    (a) Inelastically scattered intensity as a function of time for $\Omega_\mathrm{L}=25\Gamma$ and $\Delta=0$ shown for two different optical depths $b_0$, showing the shift in Rabi frequency for higher optical depth. (b) Corresponding Fourier transforms of the signals in (a), with the vertical dashed-dotted line indicating the peak oscillation frequency obtained from Lorentzian fits.}
    \label{fig:intenfft}
\end{figure}

Since we are interested in the strong-drive regime $\Omega_\mathrm{L}\gg\sqrt{\Gamma^2+4\Delta^2}$, we first focus on the inelastically scattered light $I^\mathrm{inel}$, while the contribution from elastic scattering is discussed later. At resonance ($\Delta=0$), when driven by a laser with electric field amplitude $E$, independent atoms oscillate between the ground and excited state with the Rabi frequency $\Omega_\mathrm{L}=dE/\hbar$ ($d$ the electric dipole transition moment and $\hbar$ Planck's constant). As the atomic populations start oscillating, the intensity also displays these Rabi oscillations, see Fig.~\ref{fig:intenfft}(a). For an optically dilute cloud, ($b_0=1$, red solid curve), where the interactions are very weak, the oscillations are essentially those of a single atom. In particular, their frequency is given by the (single-atom) generalized Rabi frequency, $\Omega_\mathrm{G}^{(1)}=\sqrt{\Omega_\mathrm{L}^2+\Delta^2}$. This is confirmed by analyzing the oscillations in frequency space, where the spectrum is centered around $\Omega_\mathrm{G}^{(1)}$, see Fig.~\ref{fig:intenfft}(b). 

In contrast, for an optically dense cloud we observe a clear modification of the oscillation frequency, of the order of a fraction of $\Gamma$, see Fig.~\ref{fig:intenfft} (blue dashed curve for $b_0=81$). As the interaction between the atoms displaces the effective resonance frequency by $\delta_\mathrm{c}$, the detuning of the driving laser gets modified as $\Delta\to \Delta-\delta_\mathrm{c}$ and the oscillation frequency becomes $\Omega^{(N)}_{\mathrm{G}} = \sqrt{\Omega_\mathrm{L}^2 + (\Delta - \delta_\mathrm{c})^2}$. We also note that the damping of the oscillations increases with the interaction (or, equivalently, the width of the peak in Fig.~\ref{fig:intenfft}(b)), an effect previously associated with collective radiative decay (superradiance)~\cite{EspiritoSanto2020}.

\section{Attenuation effects}

We here highlight that the average field reaching the atoms in an optically thick cloud is actually reduced, as compared to the single-atom one, corresponding to the incident laser field. As the probe propagates through the medium, scattering events attenuate its intensity, modifying the effective Rabi frequency seen by atoms deeper in the cloud. In this section we verify that this attenuation in our model follows the predictions of light transport in saturated atomic media~\cite{Reinaudi2007}, so even without a resonance displacement the attenuation of the beam modifies the average Rabi frequency over the cloud. 

The propagation of resonant light through an absorbing medium is governed by 
the Beer–Lambert law, which is nevertheless modified in the case of a strong drive to account for the intensity-dependent atoms saturation~\cite{Reinaudi2007}:
\begin{equation}
\frac{dI(z)}{dz} = -\rho(z) \sigma_{sc} 
\frac{I(z)}{1 + I(z)/I_{\mathrm{sat}}},
\end{equation}
where $I(z)$ is the transmitted intensity, $\rho(z)$ the local atomic density, $\sigma_{sc} = \lambda^2/\pi$ the scattering cross section in the scalar regime, and $I_{\mathrm{sat}}$ the saturation intensity. 

\begin{figure}[htpb]
    \centering
    \includegraphics[width=\columnwidth]{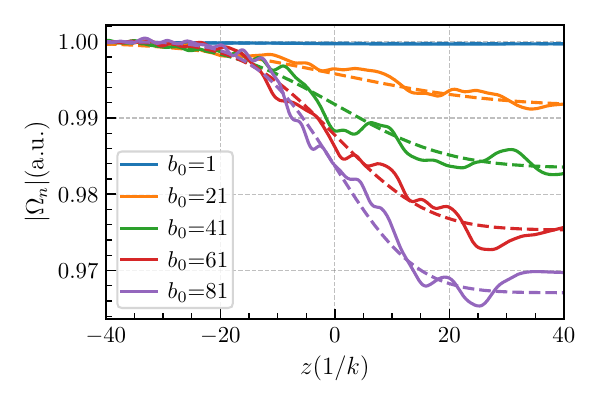}
    \caption{Local amplitude of the total field in the cloud along the laser propagation direction (solid lines), compared to Beer–Lambert predictions (dashed lines) for different optical depths $b_0$ in the steady-state. The $y$-axis is normalized to the probe’s incident Rabi frequency. The intensity is computed at the center of the cloud ($x=y=0$), at resonance $\Delta = 0$ and with $\Omega_\mathrm{L} = 25\Gamma$.}
    \label{fig:shadow}
\end{figure}

In the weak-drive (linear optics) limit $I \ll I_{\mathrm{sat}}$, this reduces to the standard exponential attenuation:
\begin{equation}
I(z) = I_0 \exp\left(-\int_0^z \rho(z') \sigma_{sc} \, dz'\right),
\end{equation}
with $b_0=\int_0^z \rho(z') \sigma_{sc} \, dz'$ the optical depth. In contrast, in the strongly driven regime, the cloud becomes     partially transparent, and the transmitted intensity follows a linear profile:
\begin{equation}
I(z) \approx I_0 - I_{\mathrm{sat}} \int_0^z \rho(z') \sigma_{sc} \, dz'.
\end{equation}
To evaluate how this attenuation manifests in our system, we analyze the amplitude of the total field $\Omega_n$ along the laser propagation axis ($x = y = 0$) in the stationary regime, where the atomic density profile is $\rho(z) = \rho_0 \exp(-z^2/2R^2)$, with 
$\rho_0 = N/(\sqrt{2\pi} R)^3$. Similarly to Eq.~\eqref{eq:localfieldWn}, the local total field is computed as
\begin{equation}
\Omega_n(\mathbf{r}) = \frac{\Omega_\mathrm{L} e^{ikz}}{2} - i \sum_n \frac{\Gamma}{2} 
\frac{e^{ik|\mathbf{r}-\mathbf{r}_n|}}{ik|\mathbf{r}-\mathbf{r}_n|} \beta_n.
\end{equation}

Figure~\ref{fig:shadow} compares the spatial dependence of the total field amplitude extracted from mean-field simulations (solid curves) with predictions from the Beer–Lambert law (dashed curves). The good agreement confirms that the attenuation of the driving field in the steady-state is well captured by the mean-field model.

This attenuation reflects the fact that the coupled dipole model~\eqref{eq:DDHamiltonian} describes the interference of the driving field with the one generated by the excited atoms, which results in a screening of this drive. Nevertheless, monitoring the Rabi oscillation frequency $\Omega_\mathrm{G}^{(N)}$ by sweeping the laser frequency over the atomic resonance allows to overcome this effect, since the attenuation is expected to not depend on the sign of the detuning.

\section{Detuning Sweep and effective scaling of the resonance displacement}

\begin{figure}[htpb]
    \centering
    \includegraphics[width=\columnwidth]{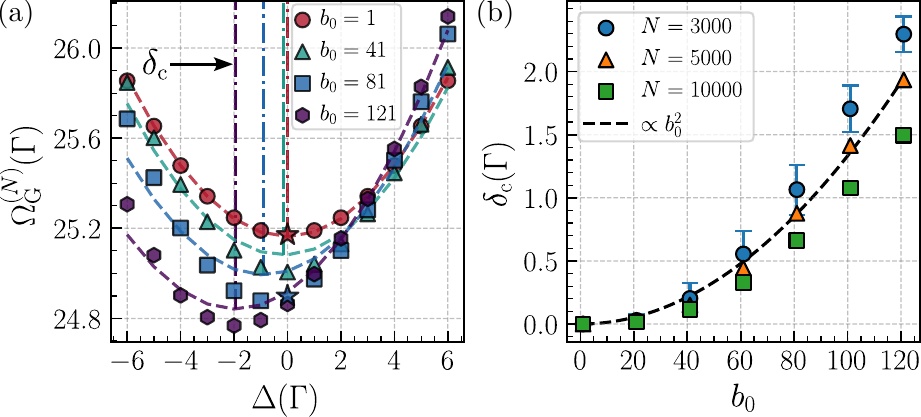}
    \caption{(a) Extracted oscillation frequency $\Omega^{(N)}_{\mathrm{G}}$ as a function of the detuning $\Delta$ for different $b_0$, with dashed lines showing fits to Eq.~\eqref{eq:Omegacol}. The two star markers correspond to the specific curves in Fig.~\ref{fig:intenfft}. (d) Collective resonance displacement $\delta_\mathrm{c}$ extracted from the fits in (b), as a function of the optical depth and for different atom numbers. Despite minor deviations, the results confirm that optical depth is the relevant control parameter for the resonance displacement. The black dashed curve corresponds to a quadratic scaling.}
    \label{fig:Swipe}
\end{figure}

In Fig.~\ref{fig:Swipe}(a), such frequency detuning scans reveal two effects. First, an overall reduction of the frequency for increasing $b_0$ occurs (curves from top to bottom), which can be attributed to the attenuation through the thick cloud discussed above. Second, a displacement of the resonance is observed, which corresponds to the minimum oscillation frequency (see vertical dash-dotted curves). This second effect is a collective phenomenon since it arises with the optical depth $b_0$, whereas the attenuation effect leads to a modification of the Rabi frequency which is expected to remain symmetric around $\Delta=0$.

This displacement $\delta_\mathrm{c}$ modifies the transition frequency for the cloud as $\omega_\mathrm{a}\to\omega_\mathrm{a}+\delta_\mathrm{c}$, which modifies the generalized Rabi frequency as follows:
\begin{equation}
\Omega_\mathrm{G}^{(N)}=\sqrt{\overline{\Omega}^2+(\Delta-\delta_\mathrm{c})^2},
\label{eq:Omegacol}
\end{equation}
with $\overline{\Omega}$ the laser Rabi frequency modified by the attenuation. The minimum frequency of the Rabi oscillations is thus reached at $\Delta=\delta_\mathrm{c}$. In our analysis, the evolution of $\Omega_\mathrm{G}^{(N)}$ is fitted using Eq.~\eqref{eq:Omegacol}, with $\overline{\Omega}$ and $\delta_\mathrm{c}$ as free parameters.

\begin{figure*}[htpb]
    \centering
    \includegraphics[width=0.9\linewidth]{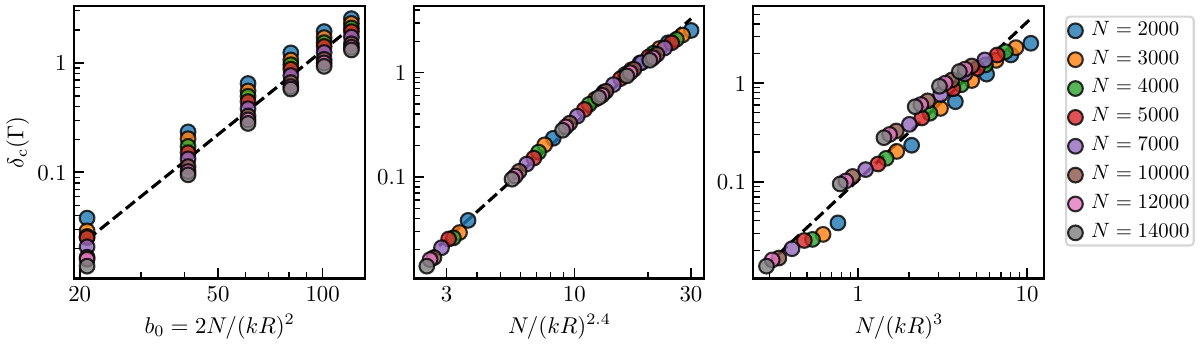}
    \caption{Collective resonance displacement $\delta_\mathrm{c}$ as a function of (a) optical depth $b_0 = 2N/(kR)^2$, (b) rescaled parameter $N/(kR)^{2.4}$, and (c) normalized atomic density $N/(kR)^3$. Each color represents a different atom number. The best data collapse occurs for $\alpha = 2.4$, suggesting that the collective displacement scales primarily with optical depth, with a minor correction related to spatial density.}
    \label{fig:SM2}
\end{figure*}

The collective nature of the displacement is first confirmed by monitoring its growth with the optical depth, see Fig.~\ref{fig:Swipe}(b). In this plot we observe a clear trend of the resonance displacement $\delta_\mathrm{c}$ with the optical depth, extracted from the minima of the detuning sweep curves. However, simulating the intensity oscillations for clouds with different atom numbers and sizes reveals a noticeable spread between the curves even for the same values of optical depth. This indicates that other effects are into play, and to identify a more accurate scaling variable, we explored a family of parameters of the form $N/(kR)^\alpha$, aiming to collapse all data onto a single curve. As shown in Fig.~\ref{fig:SM2}, the best collapse is obtained for $\alpha = 2.4$, determined by maximizing the coefficient of determination $R^2$ from a linear fit of $\log \delta_\mathrm{c}$ vs. $\log(N/(kR)^\alpha)$. This scaling deviates slightly from the conventional optical depth $\alpha = 2$, but is much closer to it than to a volume-density scaling ($\alpha = 3$), suggesting that the collective resonance displacement is governed primarily by the optical depth, with a weak residual dependence on spatial density or geometry. This scaling behavior supports the interpretation that the observed displacement arises from long-range dipole-dipole interactions, akin to the reports of cooperative phenomena such as superradiance and subradiance in dilute clouds~\cite{Guerin2016, Araujo2016,Guerin2017}.

While the densities considered in this work are substantial due to the limited number of atoms which can be simulated, we find that the density effects affect only weakly the dynamics. Near-field contributions, which could become significant at high density, primarily affect the optical-coherence component~\cite{Moreira2024} and thus contribute weakly in the strong-driving regime considered here. Moreover, in a dilute disordered clouds, polarization-dependent near-field interactions are typically negligible. This justifies the choice of the scalar model for the light, which allows us to study the response of disordered systems under strong driving, even if numerically the densities are not negligible. In particular, the slight deviation from the $b_0^2$ scaling suggests that optical depth alone does not fully account for the effect, and that simulations at larger atom numbers may help resolve the residual influence of spatial density. Reaching this limit, however, requires simulating significantly larger systems, which remains computationally demanding due to the nonlinear nature of the coupled equations.

\section{A collective dynamical resonance displacement} 

\begin{figure}[htbp]
    \centering
        \includegraphics[width=\columnwidth]{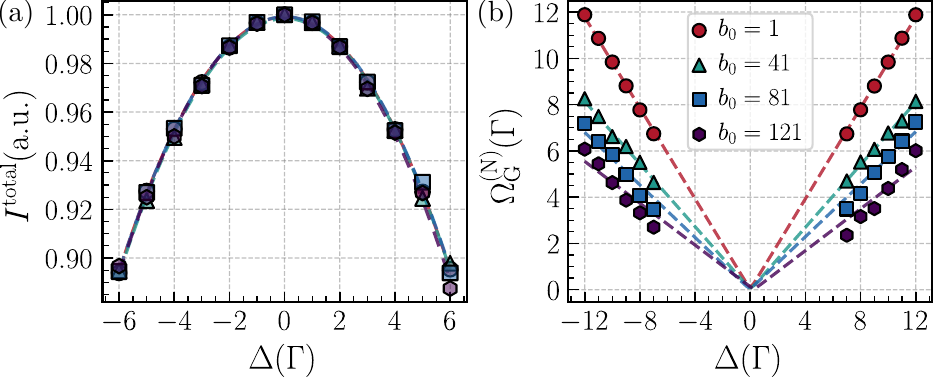}
    \caption{(a) Normalized total steady-state scattered intensity at an angle $\theta=47^\circ$ as a function of $\Delta$, for $\Omega_\mathrm{L}=25\Gamma$ and different optical depths $b_0$. No displacement in the resonance peak is observed, highlighting the dynamical nature of the reported effect. (b) Oscillation frequency $\Omega^{(N)}_{\mathrm{G}}$ for $\Omega_\mathrm{L}=0.1\Gamma$, corresponding to the linear regime ($s\ll1$), extracted using the detuning sweep method from total scattered intensity. No significant displacement is observed with increasing optical depth, showing that the resonance displacement emerges only in the saturated regime. The system is not simulated for small detuning values as for low $|\Omega_\mathrm{L}|$ and $|\Delta|$ the oscillations are overdamped and it is not possible to extract the Rabi frequency.}
    \label{fig:SSandLinear}
\end{figure}

Frequency displacements due to dipole-dipole interactions have been predicted in several configurations, especially in the linear ``single-photon'' regime where the atoms can be described as classical oscillators~\cite{Svidzinsky2010,Cottier2018}. In particular, monitoring the transmission as the laser frequency is tuned also lead to reports of frequency shifts in the steady state~\cite{Jenkins2016,Bettles2020,Friedberg2022,hofer2024}, with the inhomogeneous broadening playing an important role~\cite{Javanainen2014}. We point out that in the steady state ($t\gg 1/\Gamma$), no shift is present in our saturated atomic cloud as the total scattered intensity is maximum when the laser frequency is tuned at the atomic resonance, see Fig.~\ref{fig:SSandLinear}(a). Hence the modification reported here has a \textit{dynamical} nature.

Note that a collective modification of the oscillations frequency of the intensity can also be observed in the linear regime, using a weak-drive switch-on protocol. Reported in Ref.~\cite{Guerin2019} and described through the fully-classical approach of coupled dipoles~\cite{Guerin2019,EspiritoSanto2020}, it is associated to a splitting of the resonant frequency, and it is a free-space analog of the \textit{mode splitting} commonly observed in optical cavities~\cite{Sanchez1983,Thompson1992,Yoshie2004,Colombe2007}. We have here checked that the semi-classical approach~(\ref{eq:bn}--\ref{eq:zn}) captures well the splitting of the linear regime. In this case, the spectrum is fitted as $\Omega^{(N)}_{\mathrm{G}} = \alpha \sqrt{\Omega_\mathrm{L}^2 + (\Delta - \delta_\mathrm{c})^2}$, with $\Omega_\mathrm{L}\ll\Gamma$ , and the $\alpha$ accounting for the reduction in the oscillation frequency reported in Ref.~\cite{Guerin2019}. As illustrated in Fig.~\ref{fig:SSandLinear}(b) for $\Omega_\mathrm{L}=\Gamma/10$, the oscillation frequency modification induced by this splitting is symmetric around the atomic resonance, so it does not correspond to a displacement of the atomic resonance -- note that in this regime the total intensity is considered, and it is dominated by the elastic component $I^\mathrm{el}$, calculated at a fixed polar angle $\theta=47^\circ$ and averaged over azimuthal angles. Hence, the dynamical resonance displacement observed in the strong drive regime is thus specific to Rabi oscillations of two-level emitters ($\Omega_\mathrm{L}\gg\Gamma$) rather than of coupled classical oscillators ($\Omega_\mathrm{L}\ll\Gamma,\,\Delta$).

\section{Impact of the elastically scattered light} 
\begin{figure}[htbp]
    \centering
        \includegraphics[width=\columnwidth]{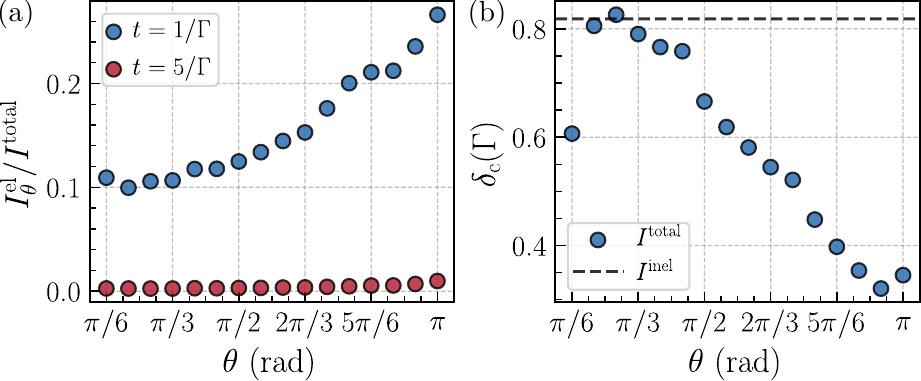}
    \caption{(a) Ratio of elastically scattered intensity, as a function of the scattering angle $\theta$, averaged over the azimuthal angle, at time $t=1/\Gamma$ and $5/\Gamma$ for a cloud of optical depth $b_0=81$. The blue dots illustrate the strong angular dependence of the elastic scattering at early times, while the red dots show how it collapses at later times. (b) Dynamical resonance displacement $\delta_\mathrm{c}$ for a cloud of optical depth $b_0=81$ as a function of the angle for the total light scattered (blue dots), which presents a strong dependence on the angle of observation $\theta$; the black dashed corresponds to the displacement for the inelastic light only, and it does not present any angular dependence.}
    \label{fig:elastic}
\end{figure}

The dynamic results presented in Fig.~\ref{fig:intenfft} were computed considering the inelastically scattered light only, motivated by the limit $\Omega_\mathrm{L}\gg\sqrt{\Gamma^2+4\Delta^2}$. While in the mean-field approximation the component $I^\mathrm{inel}$ does not present any angular dependence, the elastic component does. Furthermore, shortly after the laser is switched on, the contribution of elastic scattering is much larger than in the steady-state, and it is particularly strong in the backward direction. This is illustrated in Fig.~\ref{fig:elastic}(a), where the ratio of elastic to total scattered light is plotted as a function of the angle from the pump axis at $t=1/\Gamma$ -- elastic scattering exhibits an interference pattern, thus bringing a spatial dependence to the scattered intensity. In the backward direction, elastic represents almost one fourth of the total scattered light, despite a driving Rabi frequency $\Omega_\mathrm{L}=25\Gamma$ (in the steady state the ratio is $\approx 0.1\%$ for independent atoms). Indeed, at initial time the atoms are not yet saturated, and strong atomic coherences build up due to the coherent drive.

As a consequence, monitoring the \textit{total} light scattered by the cloud, rather than the inelastic component only, leads to an oscillating intensity in space, and thus a displacement that depends on the angle of observation~\cite{Hsu2024}. We present this feature in Fig.~\ref{fig:elastic}(b), where a substantial reduction of the resonance displacement is observed as the total light is monitored toward the backward direction. In experiments, the resonance displacement in the inelastic contribution could be isolated, for example, by filtering out the elastic component.

Finally, let us comment on the scaling of the dynamical resonance displacement. First, a reduction is observed as the laser Rabi frequency $\Omega_\mathrm{L}$ is increased, see Fig.~\ref{fig:schemeAndResult}(b). This effect can be attributed to the fact that the interactions ($\hat{\sigma}^+_n\hat{\sigma}^-_m$ exchange terms) are based on the atomic dipole moments, whose expected value ($\langle\hat{\sigma}^+_n\rangle$) decreases with the drive strength $\Omega_\mathrm{L}$. Another perspective on this effect is that as the Rabi frequency $\Omega_\mathrm{L}$ on atom $n$ is increased, the relative weight of the radiation from the other atoms becomes weaker ($\sum_{m\neq n}G^s_{nm}\langle\hat{\sigma}^-_m\rangle$ is bounded, while $\Omega_\mathrm{L}$ is not). In other terms, the bleaching of the atoms by the strong laser light reduces the contribution from (dipole-dipole) cooperative effects, and effectively reduces the observed resonance displacement.


\section{Conclusions} We have identified a collective modification in the Rabi oscillation frequency of optically thick atomic clouds under strong drive, arising from a collective atomic resonance displacement due to dipole-dipole interactions. This displacement emerges as the system is suddenly driven far from the ground state and vanishes in the steady-state, confirming its dynamical nature. Our simulations show that the resonance displacement scales quadratically with the optical depth and decreases with increasing saturation, reflecting the interplay between dipole-dipole interactions and population redistribution. Importantly, this effect presents distinct features from collective shifts previously reported in the weak-drive (``linear'' or ``single-photon'') regime, and also cannot be explained by attenuation effects alone, demonstrating that cooperative phenomena remain relevant in saturated clouds with many excitations.

Throughout this work we focus on dilute disordered clouds, in which the long-range dipole–dipole term dominates. This approximation is particularly relevant for clouds in magneto-optical traps. We note, however, that the long-range mechanism responsible for the dynamical shift can also be dominant in metrological platforms, such as optical lattice clocks operated near the magic angle, where the near-field terms vanish. Another approximation done is neglecting the atoms displacement and Doppler broadening, assuming that the oscillations occur on timescales shorter than the ones associated to motional broadening. This sets an upper temperature below which the collective oscillations are expected to be visible.

Finally, the choice of the driving strength is an important element for observing the effect. Very strong driving improves the visibility of Rabi oscillations, but at the same time it reduces the relative weight of dipole--dipole interactions, and thus the magnitude of the resonance displacement. While our analysis did not identify a unique optimal value for experimental observation, the drive strengths used in this work correspond to a compromise between oscillation contrast and a detectable displacement.

More broadly, this work highlights a new regime of cooperative dynamics in light-matter interaction, where saturation does not suppress but rather reshapes collective behavior. This can be particularly important when dynamical many-body effects are probed and the cloud holds a large number of excitations. In particular, in the context of metrology and Ramsey spectroscopy, the area of short pulses may be affected by such collective effects, despite a Rabi frequency much larger than the transition linewidth. Future investigations could explore beyond-mean-field effects, while experimentally a similar setup and procedure can be implemented for cold atoms in a magneto-optical trap, or analyzed in optical atomic clocks.



\begin{acknowledgments}
\section{Acknowledgments}
M.\,A.\,F.\,B., and R.\,B. acknowledge the financial support of the São Paulo Research Foundation
(FAPESP) (Grants No. 2021/02673-9, 2022/00209-6, 2023/15457-8 and 2023/03300-7) and from the Brazilian CNPq (Conselho Nacional de Desenvolvimento Científico e Tecnológico), Grant No. 313632/2023-5. The authors acknowledge the financial support of CAPES-COFECUB (CAPES, Grant No. 88887.711967/2022-00 and COFECUB, Grant No. Ph 997/23). M.\,H. and R.\,B. have been supported by the the UCA J.E.D.I. Investments (ANR-15-IDEX-01).
\end{acknowledgments}

\bibliography{Biblio}
\onecolumngrid
\newpage

\end{document}